# Bistochastic privacy

Nicolas Ruiz and Josep Domingo-Ferrer

Universitat Rovira i Virgili
Departament d'Enginyeria Informàtica i Matemàtiques
Av. Països Catalans 26, 43007 Tarragona, Catalonia
`{nicolas.ruiz,josep.domingo}@urv.cat`

**Abstract.** We introduce a new privacy model relying on bistochastic matrices, that is, matrices whose components are nonnegative and sum to 1 both row-wise and column-wise. This class of matrices is used to both define privacy guarantees and a tool to apply protection on a data set. The bistochasticity assumption happens to connect several fields of the privacy literature, including the two most popular models, *k*-anonymity and differential privacy. Moreover, it establishes a bridge with information theory, which simplifies the thorny issue of evaluating the utility of a protected data set. Bistochastic privacy also clarifies the trade-off between protection and utility by using bits, which can be viewed as a natural currency to comprehend and operationalize this trade-off, in the same way than bits are used in information theory to capture uncertainty. A discussion on the suitable parameterization of bistochastic matrices to achieve the privacy guarantees of this new model is also provided.

## 1 Introduction

In the clash between pervasive big data collection and exploratory big data analytics on the one hand, and stronger data protection legislation on the other hand, anonymization stands out as a way to reconcile both sides. Indeed, the European General Data Protection Regulation (GDPR, [8]), which can be viewed as an epitome of strong regulation, establishes that personally identifiable information (PII) is no longer personal after anonymization. Hence, anonymized data fall outside the scope of privacy regulations and can be freely stored and processed. For anonymization to provide effective privacy protection, it has to prevent disclosure. Disclosure can occur if an intruder can determine the identity of the subject to whom a piece of anonymized data corresponds —re-identification disclosure—, or can estimate the value of a subject's confidential attribute after seeing the anonymized data —attribute disclosure.

The traditional approach to anonymization, still very dominant among statistical agencies, can be called utility-first. It essentially consists of leveraging a repertoire of masking methods collectively known as statistical disclosure control (SDC, [9]). An SDC method with a heuristic parameter choice and suitable utility preservation properties is run to anonymize the original data. Then the risk of disclosure is assessed empirically (for example using record linkage between the original and the anonymized data) or analytically (using generic measures or measures tailored to a specific SDC method). If the remaining risk is deemed too high, the data protector tries an SDC method having more privacy-stringent parameters and generally more utility loss. This process is iterated until the risk is low enough.

The computer science approach to anonymization could be termed privacy-first, and it is based on privacy models. A privacy model is a privacy condition dependent on a parameter that guarantees an upper bound on the re-identification risk and perhaps on the attribute disclosure risk. Each privacy model can be enforced using one or several SDC methods. There are currently two main families of privacy models, one based on $k$-anonymity [14] and the other on $\varepsilon$-differential privacy [7]. As shown in [2], the two families are complementary and have their own merits.

A problem with the current state of the art in the literature is that it appears as a variegated collection of SDC methods and privacy models. Whereas the permutation model [4] has been proposed to give a conceptual connection among SDC methods, no encompassing framework exists for privacy models. The ambition of this paper is to break ground towards a framework that not only unifies the two main families of privacy models —differential privacy and k -anonymity— but also aligns anonymization with information theory, which in turn simplifies what is meant by utility for an anonymized data set. We introduce bistochastic privacy, a specific form of randomized response in which the anonymized data $Y$ are obtained from the original data $X$ using Markov transition matrices that are bistochastic, that is, whose components are nonnegative and sum to 1 both row-wise and column-wise.

Section 2 connects bistochastic matrices with differential privacy, $k$-anonymity and SDC. A new privacy model, aligning information theory and privacy is then presented in Section 3, while Section 4 discusses the parametrization of bistochastic matrices. Finally, conclusions and directions for future work are gathered in Section 5. The Appendix gives background on randomized response, the permutation model of SDC and information theory.

## 2   Connections between SDC, differential privacy and *k*-anonymity through bistochastic matrices

To the best of our knowledge, this is the first time that bistochastic matrices are explicitly considered in the privacy literature. However, it happens that, without it being clearly stated, they have already been implicitly used. In what follows, we establish novel theoretical results showing that the bistochasticity assumption is a connector across SDC, differential privacy and *k*-anonymity.

## 2.1 Connection with SDC

We will assume a randomized response matrix *P* (see Expression (A.1) in the appendix) that fulfills the additional left stochasticity constraints that $\sum_{u=1}^{r} p_{uv} = 1 \ \forall v = 1, \ldots, r$. This makes *P* bistochastic (left stochasticity implies that any anonymized categories must come from the original categories). The following result then follows:

***Theorem 1 (Birkhoff-Von Neumann [12])***: *If an r×r matrix P is bistochastic, then there exist $\lambda_1, \ldots, \lambda_J \geq 0$ with $\sum_{j=1}^{J} \lambda_j = 1$ and $P_1, \ldots, P_J$ permutation matrices such that:*

$$P = \sum_{j=1}^{J} \lambda_j P_j \quad (1)$$

Theorem 1 states that any bistochastic matrix can always be expressed as a convex combination of permutation matrices. Note that while there are *r!* possible permutations of *r* categories, every *r* × *r* doubly stochastic matrix can be represented as a convex combination, which may not be unique, of at most *r²−2r+2* permutation matrices [12].

This result directly establishes a connection with SDC through the permutation model. In fact, SDC can be viewed as a specific case of a more general approach that uses bistochastic matrices to perform anonymization. The permutation model considers a crisp permutation within the data set domain: it yields values occurring in the data set, except perhaps for a small noise addition that does not alter ranks. In contrast, a bistochastic matrix is described by Theorem 1 as a probabilistic model of permutation within the domain of attributes:

- The bistochastic transition matrix maps true values in the original data set to reported values that can in general be any value in the domain of the attributes —perhaps very different from the attribute values occurring in the data set.
- Expression (1) can be viewed as a probabilistic permutation: each permutation matrix $P_j$ has a probability $\lambda_j$ of being actually used. Only if $\lambda_j = 1$ for some *j* in Expression (1), which describes the functioning of any SDC methods, is permutation $P_j$ certain to occur.

## 2.2 Connection with differential privacy

Differential privacy (DP) is a privacy model that can be enforced using a variety of SDC techniques [13]. In what follows, we choose Randomized Response (RR) as a technique to enforce DP. This is a legitimate setting, as during the inception of differential privacy, randomized response was already considered as a method to produce differentially private data sets. Thus, the connection established can be viewed as reasonably general. It follows that differential privacy constraints on an RR scheme happen to enforce bistochasticity, as is shown by the following proposition, proven in [16]:

**Proposition 1**: The r×r matrix P of an ε-differentially private randomized response scheme is of the form:

$$p_{uv} = \begin{cases} \frac{e^\varepsilon}{r-1+e^\varepsilon} & if\ u = v \\ \frac{1}{r-1+e^\varepsilon} & if\ u \neq v \end{cases} \quad with\ \varepsilon \geq \ln \max_{u=1,\dots,r} \frac{\max_{u=1,\dots,K} p_{uv}}{\min_{u=1,\dots,K} p_{uv}} \quad (2)$$

Expression (2) describes a bistochastic matrix, as both its rows and columns sum to 1. Note also that taking $\varepsilon = 0$ in this matrix yields perfect secrecy (see Appendix), as the probabilities within each column are identical.

More generally, this result sheds an alternative light on the functioning of differential privacy, at least when it is attained through RR. To see this, assume $r=3$. In the extreme case of the strictest differential privacy, i.e. when $\varepsilon = 0$, Expression (2) implies that all components of *P* must be equal to 1/3. Following Theorem 1, the associated differentially private randomized response scheme can be expressed as the following combination of permutation matrices:

$$P = \begin{pmatrix} 1/3 & 1/3 & 1/3 \\ 1/3 & 1/3 & 1/3 \\ 1/3 & 1/3 & 1/3 \end{pmatrix} = \frac{1}{3}\begin{pmatrix} 1 & 0 & 0 \\ 0 & 1 & 0 \\ 0 & 0 & 1 \end{pmatrix} + \frac{1}{3}\begin{pmatrix} 0 & 0 & 1 \\ 1 & 0 & 0 \\ 0 & 1 & 0 \end{pmatrix} + \frac{1}{3}\begin{pmatrix} 0 & 1 & 0 \\ 0 & 0 & 1 \\ 1 & 0 & 0 \end{pmatrix}.$$

Clearly, with the strictest setting, no permutation pattern is favored. However, for $\varepsilon = 2$, one gets:

$$P = \begin{pmatrix} \frac{e^2}{2+e^2} & \frac{1}{2+e^2} & \frac{1}{2+e^2} \\ \frac{1}{2+e^2} & \frac{e^2}{2+e^2} & \frac{1}{2+e^2} \\ \frac{1}{2+e^2} & \frac{1}{2+e^2} & \frac{e^2}{2+e^2} \end{pmatrix}$$

$$= \frac{e^2}{2+e^2}\begin{pmatrix} 1 & 0 & 0 \\ 0 & 1 & 0 \\ 0 & 0 & 1 \end{pmatrix} + \frac{1}{2+e^2}\begin{pmatrix} 0 & 0 & 1 \\ 1 & 0 & 0 \\ 0 & 1 & 0 \end{pmatrix} + \frac{1}{2+e^2}\begin{pmatrix} 0 & 1 & 0 \\ 0 & 0 & 1 \\ 1 & 0 & 0 \end{pmatrix}.$$

When the constraints imposed by differential privacy are relaxed, the probability of not altering the data (the identity matrix being a special case of permutation) is favored with a probability of $\frac{e^2}{2+e^2} = 0.78$, while other permutation patterns have a probability of 0.11 of being taken.

The usual notion of differential privacy is that the presence or absence of any given record in a data set cannot be noticed, up to *exp(ε)*, upon seeing anonymized outputs on the data set. When differential privacy is achieved via RR and is viewed through the lens of bistochastic matrix, it can be seen as ensuring blindness on how attribute categories are permuted. The strictest enforcement of differential privacy ($\varepsilon = 0$) amounts to random permutation and, as we saw, to perfect secrecy. With a laxer enforcement, some specific permutation patterns are more likely to occur. In Expression (2) we see that for $\varepsilon = 2$ not enough privacy is provided, because the chances of releasing the original data unaltered are 78%. Thus, *the privacy budget ε can also be seen as being proportional to the probability of not permuting the data*. Hence, too large a budget does not provide sufficient deniability. Conversely, the smaller the budget, the more credible is an individual who can deny that her reported category is her original category. Therefore, the smaller ε, the higher is plausible deniability.

### 2.3 Connection with *k*-anonymity

A bistochastic matrix can also be parametrized to fulfill *k*-anonymity, more specifically its Anatomy variant [18]. Like standard *k*-anonymity, Anatomy relies on splitting the records in the data set into classes of at least *k* records. However, unlike standard *k*-anonymity, the quasi-identifier values within each class are not made equal. Instead, two tables are released for each class: one contains the projection of the original records of the class on the quasi-identifier attributes, and the other the projections of the original records on the rest of attributes. The correspondence between entries in the two tables of each class is not revealed: thus, if the class contains *k* records, there are *k!* possible bijections between its quasi-identifier value combinations and its value combinations for the other attributes. In particular, given a quasi-identifier combination, the probability that an intruder finds the matching confidential attribute values is at most *1/k*, as in standard *k*-anonymity (note here that *l*-diversity is not guaranteed on the rest of attributes).

Let *X* be an original data set that is "anatomized" as follows:
- Compute *k*-anonymous classes of the records. Let the number of resulting classes be *L* and the number of different quasi-identifier combinations in the *l*-th class be $n_l$, for *l = 1,...,L*.
- For each class release two tables as in Anatomy, one table containing a random permutation of quasi-identifier combinations and the other table the projections of the records on the remaining attributes (those that are not quasi-identifiers). The set of the two tables for every class constitutes the anatomized data set *Y*.

The quasi-identifier tables of the anatomized *k*-anonymous data set *Y* can be viewed as having been obtained using the following transition matrix:

$$Q = \begin{pmatrix} Q_1 & 0 & \cdots & \cdots & 0 \\ 0 & Q_2 & 0 & \cdots & 0 \\ \vdots & 0 & \ddots & \ddots & \vdots \\ \vdots & \vdots & \ddots & \ddots & 0 \\ 0 & 0 & \cdots & 0 & Q_L \end{pmatrix} \quad (3)$$

with $Q_l$ being the following $n_l \times n_l$ submatrix, for *l = 1,...,L*:

$$Q_l = \begin{pmatrix} 1/n_l & \cdots & 1/n_l \\ \vdots & \ddots & \vdots \\ 1/n_l & \cdots & 1/n_l \end{pmatrix}.$$

Each submatrix $Q_l$ randomly permutes the quasi-identifier combinations within a class. If a combination of quasi-identifiers is repeated in two different classes *i* and *j*, it is permuted differently in each class, according to the respective submatrices $Q_i$ and $Q_j$. That is, the combination has two different rows and two different columns in *Q*, specifically one row and one column in $Q_i$ and one row and one column in $Q_j$. Finally, note that $Q_l$ is bistochastic $\forall l = 1,...,L$, and that the overall *Q* is also bistochastic.

Thus, *k*-anonymity can be viewed as the application of a special parametrization of a bistochastic matrix. In fact, and as each submatrix $Q_l$ achieves perfect secrecy, *k*-anonymity can be seen as a collection of perfect privacy blocks, which is exactly the original intuition behind *k*-anonymity, gathered into a block-diagonal, bistochastic matrix.

## 3 A privacy model based on bistochastic matrices

At first sight, one could wonder about the necessity of imposing an additional constraint on RR and its ex-post version PRAM [11], some well-trodden approaches for anonymization that have proved their merits over the years. However, and beyond the appeal of the theoretical connections developed above, the interest in bistochasticity is justified by the following theorem (see Appendix A.3 for some background notion on majorization and the $\succ$ relationship):

***Theorem 2 (Hardy, Littlewood, and Polya [12]):*** *$p \succ q$ if and only if $q = P^T p$ for some bistochastic matrix P.*

Theorem 2 states that a bistochastic matrix never decreases uncertainty and is the only class of matrices to do so. In fact, and when it is not a permutation matrix, it always increase uncertainty. When *P* is only right stochastic, as in the traditional approach to RR, no particular majorization relationship emerges and the resulting anonymized attribute cannot be qualified as more (or less) uncertain (in the sense of information theory) than the original attribute. However, when *P* is bistochastic but not a permutation matrix, the anonymized attribute will always be more uncertain, i.e. it will always contain more entropy. Here lies the fundamental functioning behind the privacy model proposed in this paper. The idea is to infuse a data set with uncertainty, which in fact provides protection but at the same time degrades information.

### 3.1 Univariate bistochastic privacy

We start by the simplest case where we seek to anonymize only one attribute to prevent disclosure. In what follows, we will assume that in Expression (1) $p_{uv} > 0 \; \forall u, v$. The transition matrix *P* has only strictly positive entries, meaning that any individual in any of the *r* categories can be reported in the anonymized attribute in any other of the *r* categories. As some of the transition probabilities can be made as small as desired, this is not really binding for the validity of the anonymized attribute. However, this additional constraint makes *P* the transition matrix of an ergodic Markov chain [3]. In turn, that implies that *P* has a unique stationary distribution, which, as *P* is bistochastic, is the uniform distribution [3].

The entropy rate of *P* is then given by the standard formula:

$$H(P) = -\sum_{u,v=1}^{r} \mu_u p_{uv} \log_2 p_{uv}, \quad (4)$$

where $\mu_u$ denotes the uniform distribution, i.e. $\mu_u = 1/r$.

The entropy rate of P is the average of the entropies of each row of P. Note that, in the case of perfect secrecy where all probabilities in P are equal, that we will denote hereafter by $P^*$, we have $H(P^*) = \log_2 r$, which is the maximum achievable entropy for an $r \times r$ bistochastic matrix. The definition of bistochastic privacy then follows:

***Definition 1 (Univariate bistochastic privacy)****: The anonymized version Y of an original attribute X is β-bistochastically private for 0 ≤ β ≤ 1 if:*
    i)       $Y = P^T X$ with P bistochastic
    ii)      $\frac{H(P)}{H(P^*)} \geq \beta$.

An anonymized attribute satisfies $\beta$-bistochastic privacy if it is the product of a bistochastic matrix P and the original attribute, and if the entropy rate of P is at least $100\beta$% of the maximum achievable entropy. $H(P^*)$ represents the maximum "spending" that can be allocated to privacy, and because we defined entropy with logarithm to the base 2, this maximum amount is $\log_2 r$ bits. Thus, when $\beta = 1$, all the bits have been spent and the attribute has been infused with the maximum possible amount of uncertainty; in this case, perfect secrecy is achieved and it is clear that $Y = P^{*T} X$ returns the uniform distribution. The other extreme case $\beta = 0$ means that the attribute has been left untouched and no uncertainty has been injected, i.e. $H(P) = 0$. Thus, for $0 < \beta < 1$ there lies a continuum of cases where varying amount of uncertainty bits can be injected, which will guarantee a varying amount of protection.

Here, what we mean by protection can be illustrated by assuming that an attacker has been able to re-identify an individual through her quasi-identifiers (in whatever way those have been protected), and now wants to learn the value of her confidential attribute from the bistochastically private release of this attribute, Y. If the attribute is 1-bistochastically private, nothing can be learnt by virtue of perfect secrecy. The attacker is facing a uniform distribution and at best can only perform a random guess, and the strength of plausible deniability is maximal. An alternative way to illustrate the situation faced by an attacker is to consider the quantity $2^{H(Y)}$, which yields the number of equally probable outcome values that can be represented by Y. Since Y is the uniform distribution, this number is r. One way to think about this value is that, to learn about the value of the confidential attribute of the re-identified person, an attacker is facing an imaginary dice with r sides. The targeted individual can exactly claim that strength of plausible deniability.

In this example, the links between the confidential attribute and the quasi-identifiers have been completely broken, while the distribution of the former has been completely uniformized. Thus, information has been totally lost. In addition, and because $P^*$ is singular, an estimate about the univariate distribution cannot be retrieved through the procedure described in the Appendix on randomized response. In that case, the price to pay for perfect secrecy in terms of information is maximal.

On the other hand, when $H(P) = 0$ the original information is left untouched and the data user gets the highest possible utility from the data. Consequently, moving $\beta$ between 0 and 1 in bistochastic privacy is equivalent to operating a trade-off between information and protection. The more bits are injected in the attribute *via* a bistochastic matrix, the more information is taken away from the user and traded against protection.

Unlike other privacy models, bistochastic privacy makes the trade-off between privacy and information explicit. In fact, it can be considered as a privacy *and* utility-first approach. Moreover, it also offers the additional advantage of distorting the original information of the data *always* in the same direction. This is so because, following Theorem 2, only bistochastic matrices can increase entropy (in physics for example, it is well-known that bistochastic Markov chains are the only stochastic process satisfying the second law of thermodynamics, [10]). An additional consequence of this is that, unlike other privacy models that can be attained using several SDC techniques, bistochastic privacy must be achieved using bistochastic matrices. Whereas this might be viewed as limiting, at the same time it simplifies privacy implementation, as the same entities that are used to define the privacy guarantees of the model are also used to achieve them. In the case of differential privacy, it has been recently shown that the actual protection level offered by differentially private data sets generated through different methods can be very different, even if the same level of differential privacy guarantees is enforced at the onset [13].

By always increasing entropy, bistochastic privacy always produces anonymized data that are a coarsened version of the original data. Stated otherwise, bistochastically private data are always a compact version of the original data, where some details have been lost. This is in line with intuition, to the extent that detailed information is where privacy risks reside. A popular SDC method that also coarsens data is microaggregation, which is a common approach to achieve *k*-anonymity on a numerical attribute [9]. Microaggregation reports the centroids of clusters instead of individual values. It can be noted that if the matrix of Expression (3) is applied to a numerical attribute instead of a categorical one, the product of this matrix and the original attribute will produce a microaggregated version of the latter, with the centroids being the means of the clusters.

The fact of always coarsening data means that the evaluation of information loss for bistochastically private data is simplified as it can be systematically assessed trough this lens: any analytical needs to be performed on the data can be gauged through their behavior when data are coarsened. For example, the properties of standard econometric estimators on coarsened data are already established [17]. We believe this presents a clear advantage over other privacy models and SDC methods, for which the direction in which information is distorted is often unclear, and where one must rely on specific information loss metrics related to the analytical task to be performed.

### 3.2 Bistochastic privacy at the data set level

We now consider the case of several attributes. First, we start by noting that, following the remark on microaggregation just above, bistochastic matrices can be applied on both categorical and numerical attributes. In the categorical case, the original proportions of respondents whose values fall in each of the $r$ categories will be changed, which will coarsen the distribution to deliver randomized proportions closer to the uniform distribution. In the latter case, it will tend to average the numerical values of respondents. In fact, if PRAM is used for randomization, then in a bistochastic randomized response scheme on a numerical attribute *the individuals are used as categories*. Moreover, and because *bistochastic matrices are mean-preserving*, the anonymized numerical attribute will have the same mean as the original numerical attribute (note that this would not be possible with a non-bistochastic Markov matrix, which is generally not mean-preserving).

Bistochastic randomized response scheme on a numerical attribute can be given additional intuition by considering Expression (1). As a bistochastic matrix can always be expressed as a convex combination of permutation matrices, applying a bistochastic matrix on a numerical attribute is equivalent to permuting individuals, albeit here this is done in a probabilistic way, unlike in typical permutation/swapping SDC methods.

Finally, one can note that, in the case where matrix $P^*$ (with all probabilities in it being identical) is used, all the values of the anonymized numerical attribute $Y$ are equal to the average of the original attribute $X$. This a rather extreme case of coarsening, which makes $Y$ a $k$-anonymous version of $X$ with only one cluster.

From now on, denote by $X$ a data set comprised of $K$ attributes $X_1,...,X_K$. Based on the discussion above, we will not precise if the attributes are numerical or categorical. As a result, $n_k$ will denote either the number of categories if $X_k$ is a categorical attribute, or the number of individuals $N$ if $X_k$ is a numerical attribute. $Y$, the anonymized version of $X$, is generated by injecting entropy in each attribute $k$ through $n_k \times n_k$ bistochastic matrices $P_k$:

$$P_k = \begin{pmatrix} p_{11} & \cdots & p_{1n_k} \\ \vdots & \ddots & \vdots \\ p_{n_k 1} & \cdots & p_{n_k n_k} \end{pmatrix} \quad (5)$$

where $p_{u_k v_k} = \Pr(Y_k = v_k | X_k = u_k)$ denotes the probability that the original response (or the original individual) $u_k$ in $X_k$ is reported as $v_k$ in $Y_k$, for $u_k, v_k \in \{1,...,n_k\}$. Under this procedure, the following proposition holds:

***Proposition 2****: The maximum number of bits that can be injected into a data set $X$ is $H^*(P_1^*,...,P_K^*) = \sum_{k=1}^{K} H(P_k^*)$.*

This property stems from the fact that joint entropy is always subadditive, i.e. it always hold that $H(P_1,...,P_K) \leq \sum_{k=1}^{K} H(P_k)$ [3]. This leads to the following definition:

***Definition 2 (Conservative multivariate bistochastic privacy)****: The anonymized version $Y$ of an original data set $X$ is conservatively $\beta$-bistochastically private for $0 \leq \beta \leq 1$ if:*
  i) $Y = (P_1^T X_1,...,P_K^T X_K)$ with $P_k \forall k = 1,...,K$ bistochastic
  ii) $\frac{\sum_{k=1}^{K} H(P_k)}{\sum_{k=1}^{K} H(P_k^*)} \geq \beta$

Definition 2 has the merit of simplifying the implementation of bistochastic privacy on a whole data set. The fact that each attribute is dealt with separately keeps the computational cost relatively low [6]. Moreover, estimating the distribution of the frequencies of each attribute is easily achievable because the computational cost of inverting each bistochastic matrix is also low. However, the drawback is that, because entropy is subadditive, one injects more bits than in the case of dealing directly with the joint distribution. More protection is applied and, as result, more information is lost. In particular, the dependencies between attributes may end up getting more degraded than necessary. Unnecessary information loss is only avoided in the case where all the original attributes are independent.

A way to avoid information loss when attributes are dependent is to apply a bistochastic matrix $P_J$ directly on the joint distribution $X_J = X_1 \times ... \times X_K$. This leads to the following definition:

***Definition 3 (True multivariate bistochastic privacy)****: The anonymized version $Y_J$ of a multivariate distribution $X_J$ is $\beta$-bistochastically private for $0 \leq \beta \leq 1$ if:*
  i) $\quad Y_J = P_J^T X_J$ with $P_J$ bistochastic
  ii) $\quad \frac{H(P_J)}{H(P_J^*)} \geq \beta$

While this definition of multivariate bistochastic privacy appears in principle the most appropriate one, its computational cost may however result in practical hurdles. To perform anonymization, matrix $P_J$ may reach a very large size, in particular if the original data set contains many numerical attributes. Moreover, and while it will be still possible to retrieve an estimate of the true joint distribution using the procedure described in the Appendix on randomized response, the computational cost of inverting $P_J$ grows exponentially with the number of attributes and the presence of numerical attributes. As a result, like other privacy models, bistochastic privacy is not immune to the curse of dimensionality. For this reason, Definition 2 remains more widely applicable than Definition 3.

## 4 Parameterization of bistochastic matrices

We discuss here how to achieve bistochastic privacy by the suitable parameterizations of matrices. We saw in Section 2 two possible cases that lead to differential privacy (Expression (2)) and k-anonymity (Expression (3)) guarantees. However, beyond popular privacy models more parameterizations are possible.

We start by noting that the diagonal of a bistochastic matrix is central in any construction. Indeed, the diagonal contains the probability, for an individual or a category, that the anonymized value is the true value, meaning that the diagonal values will indicate a certain level of "truthfulness" in the anonymized data. In fact, the level of truthfulness of a bistochastic matrix is related to its singularity:

**Proposition 3**: *If for a bistochastic matrix* $P_k = \begin{pmatrix} p_{11} & \cdots & p_{1n_k} \\ \vdots & \ddots & \vdots \\ p_{n_k 1} & \cdots & p_{n_k n_k} \end{pmatrix}$
$p_{u_k u_k} > 0.5 \; \forall u_k \in \{1, ..., n_k\}$, *then $P_k$ is non-singular.*

This proposition comes from the fact that a bistochastic matrix with its diagonal values superior to 0.5 is by definition a diagonally-dominant matrix, i.e. for every row of the matrix, the magnitude of the diagonal entry in a row is larger than or equal to the sum of the magnitudes of all the other (non-diagonal) entries in that row. By the Levy–Desplanques theorem [12], such matrix is always non-singular. For anonymization (and also data utility), this means that, if a bistochastic matrix is randomizing in such a way that more than half of the time the true values are reported in the anonymized data set, then the matrix is also invertible. An estimate of the univariate distribution can then always be retrieved following the procedure outlined in the Appendix on randomized response. The setting of diagonal values is thus pivotal for parameterization but it is in no way binding. One can still set an "untruthful" matrix with very small diagonal values, albeit the non-singularity of the matrix will not always be guaranteed.

A convenient way of building a bistochastic matrix is to use a special case of Toeplitz matrices, namely a circulant matrix:

$$P_k = \begin{pmatrix} p_{11} & p_{12} & p_{13} & \cdots & p_{1n_k} \\ p_{1n_k} & p_{11} & p_{12} & \cdots & p_{1(n_k-1)} \\ p_{1(n_k-1)} & p_{1n_k} & p_{11} & \cdots & p_{1(n_k-2)} \\ \vdots & \vdots & \vdots & \vdots & \vdots \\ p_{12} & p_{13} & p_{14} & \cdots & p_{11} \end{pmatrix} \quad (6)$$

In that case, the first row of $P_k$ determines all the elements of the matrix.

Another way is to consider symmetric tridiagonal matrices $P_k$ of the following form (with $\alpha_{i-1} + \alpha_i \leq 1, \forall i \in \{1, \ldots, n_k - 2\}$):

$$\begin{pmatrix} 1-\alpha_1 & \alpha_1 & 0 & 0 & \cdots & 0 \\ \alpha_1 & 1-\alpha_1-\alpha_2 & \alpha_2 & 0 & \cdots & 0 \\ 0 & \alpha_2 & 1-\alpha_2-\alpha_3 & \alpha_3 & \cdots & 0 \\ \vdots & \vdots & \vdots & \vdots & \ddots & \vdots \\ 0 & 0 & 0 & \alpha_{n_k-1} & 1-\alpha_{n_k-1}-\alpha_{n_k} & \alpha_{n_k} \\ 0 & 0 & \cdots & \cdots & \alpha_{n_k} & 1-\alpha_{n_k} \end{pmatrix} \quad (7)$$

Remark that in Expressions (7) and (3) the matrices contain zeros and thus strictly they are not describing an ergodic process. While one can always replace the zeros by an infinitesimal term γ>0 and then adjust the other strictly positive remaining terms in order to get a strictly ergodic bistochastic matrix, a way to ease implementation is to *not* adjust the strictly positive terms to get what is called a super doubly stochastic matrix, where all rows and columns sums will be infinitesimally above one. In most cases, such matrices will behave almost like purely bistochastic ergodic matrices [10].

The latter way is the one we have followed in the examples of Table 1, where we give the number of bits of selected bistochastic matrices expressed as a percentage of the maximum possible number of bits achieved in the case of perfect secrecy, i.e. we report directly the $\beta$'s. We consider 3 parameterizations for each type of bistochastic matrix considered: *i)* differential privacy following Expression (2) for ε=5, 3 and 1, *ii)* k-anonymity following Expression (3) for k=2, 3 and 6, *iii)* a tridiagonal matrix following Expression (7) with $\alpha_{i-1}$ and $\alpha_i$= 0.1, 0.3 and 0.4, and *iv)* a circulant matrix following Expression (6) with $p_{11}$=0.9, 0.6 and 0.2 (while the remaining probabilities in each row are all equal and add to $1 - p_{11}$). The cases are set to go each time in the direction of more entropy and less truthfulness. The matrices generated are *12 × 12* in size, thus meant to be applied on a numerical attribute with 12 individuals or on a categorical attribute with 12 categories.

**Table 1.** Example of bistochastic guarantees. Each column corresponds to a different parameter value.

|  | Distance to perfect secrecy ($\beta$'s) | | |
|---|---|---|---|
| **Parametrization using:** | 1 | 2 | 3 |
| Differential privacy | 17% | 60% | 97% |
| K-anonymity | 28% | 56% | 72% |
| Tridiagonal matrix | 24% | 35% | 40% |
| Circulant matrix | 21% | 63% | 93% |

In this example, the injection of $\log_2 12 = 3.6$ bits in the attribute achieves perfect secrecy, and one can see that the strictest parameter values of differential privacy and *k*-anonymity in Table 1 come relatively close to this amount. Moreover, as differential privacy via RR gives a circulant matrix (see Expression (2)), it is not surprising that our circulant matrix parameterization happens to mimic differential privacy quite closely. While a privacy model in itself, bistochastic privacy can ease the comparison of performances across privacy models, both in terms of privacy but also of information loss, through the $\beta$'s values.

Note that to achieve bistochastic privacy, one just needs to select appropriate bistochastic matrices. To that end, the only information required on the data set to be anonymized is its size in terms of number of individuals and attributes and the number of categories for each categorical attribute. Therefore an agent, independent of the data controller, say a "data protector", can generate the appropriate matrices. The parameter $\beta$ for those matrices will depend on the environment and the desired protection-utility trade-off.

## 5    Conclusions and future research

In this paper, we have proposed bistochastic privacy, a new model that aligns privacy with information theory and unifies the main privacy models in use, in addition to connecting with the permutation model that was shown to underlie all statistical disclosure control methods [4]. The functioning of this new model also clarifies and operationalizes the trade-off between protection and utility by expressing it in terms of bits, a natural unit of privacy and information loss.

This paper opens several lines for future research. One of them is to conduct further empirical work on real-life data sets. Another is to investigate if recent solutions developed to mitigate the dimensionality problem in RR can be adapted to the present model [6]. Yet another challenge is to extend bistochastic privacy to generate new privacy models that may be more suitable for data that are unstructured or dynamic.

# Appendix

## A.1 Randomized response

Let $X$ denotes an original categorical attribute with $1, \ldots, r$ categories, and $Y$ its anonymized version. Given a value $X = u$, randomized response (RR, [1]) computes a value $Y = v$ by using an $r \times r$ Markov transition matrix:

$$P = \begin{pmatrix} p_{11} & \cdots & p_{1r} \\ \vdots & \ddots & \vdots \\ p_{r1} & \cdots & p_{rr} \end{pmatrix} \qquad (A.1)$$

where $p_{uv} = \Pr(Y = v | X = u)$ denotes the probability that the original response $u$ in $X$ is reported as $v$ in $Y$, for $u, v \in \{1, \ldots, r\}$. To be a proper Markov transition matrix, it must hold that $\sum_{v=1}^{r} p_{uv} = 1 \ \forall u = 1, \ldots, r$. $P$ is thus right stochastic, meaning that any original category must be spread along the anonymized categories.

The usual setting in RR is that each subject computes her randomized response $Y$ to be reported instead of her true response $X$. This is called the *ex-ante* or local anonymization mode. Nevertheless, it is also possible for a (trusted) data collector to gather the original responses from the subjects and randomize them in a centralized way. This *ex-post* mode corresponds to the Post-Randomization method (PRAM, [11]). Apart from who performs the anonymization, RR and PRAM operate the same way and make use of the same matrix $P$.

Let $\pi_1, \ldots, \pi_r$ be the proportions of respondents whose true values fall in each of the $r$ categories of $X$; let $\lambda_v = \sum_{u=1}^{r} p_{uv} \pi_u$ for $v=1,\ldots,r$ be the probability of the reported value $Y$ being $v$. If we define by $\lambda = (\lambda_1, \ldots, \lambda_r)^T$ and $\pi = (\pi_1, \ldots, \pi_r)^T$, then we have $\lambda = P^T \pi$. Furthermore, if $P$ is nonsingular, it is proven in [1] that an unbiased estimator $\hat{\pi}$ of $\pi$ can be obtained as $\hat{\pi} = (P^T)^{-1} \lambda$. Thus, univariate frequencies can be easily retrieved from the protected data set. Note that this procedure does not entail any privacy risk as only some estimates of the frequencies are retrieved, not specific responses that can be traced back to any individual.

RR is based on an implicit privacy guarantee called *plausible deniability* [5]. It equips the individuals with the ability to deny, with variable strength according to the parameterization of $P$, that they have reported a specific value. In fact, the more similar the probabilities in $P$, the higher the deniability. In the case where the probabilities within each column of $P$ are identical, it can be proved that *perfect secrecy* in the Shannon sense is reached [15]: observing the anonymized attribute $Y$ gives no information at all on the real value $X$. Under such configuration, a privacy breach cannot originate from the release of an anonymized data set, as the release does not bring any information that could be used for an attack. However, as exposed in the paper, the price to pay in terms of data utility is high.

## A.2 The permutation model of SDC

The permutation model of statistical disclosure control conceptually unifies SDC methods by viewing them basically as permutation [4]. Consider an original attribute $X = \{x_1, \ldots, x_n\}$ observed on $n$ individuals and its anonymized version $Y = \{y_1, \ldots, y_n\}$. Assume these attributes can be ranked —even categorical nominal attributes can be, using a semantic distance. For $i = 1$ to $n$: compute $j = Rank(y_i)$ and let $z_i = x_{(j)}$, where $x_{(j)}$ is the value of $X$ of rank $j$. Then call attribute $Z = \{z_1, ..., z_n\}$ the *reverse-mapped* version of $X$. For example, if an original value $x_1 \in X$ is anonymized as $y_1 \in Y$, and $y_1$ is, say, the 3rd smallest value in $Y$, then take $z_1$ to be the 3rd smallest value in $X$. If there are several attributes in the original data set $X$ and anonymized data set $Y$, the previous reverse-mapping procedure is conducted for each attribute; call $Z$ the data set formed by reverse-mapped attributes.

Note that: *i)* a reverse-mapped attribute $Z$ is a permutation of the corresponding original attribute $X$; *ii)* the rank order of $Z$ is the same as the rank order of $Y$. Therefore, any SDC method for microdata —individual records— is functionally equivalent to permutation —transforming data set $X$ into $Z$— followed by residual noise —transforming $Z$ into the anonymized data set $Y$. The noise added is residual because by construction the ranks of $Z$ and $Y$ are the same.

### A.3 Information theory

Classically, information theory approaches the notion of information contained in a message as capturing how much the message reduces uncertainty about something [10]. As a result, in this theory information shares the same definition as entropy and choosing which term to use depends on whether it is given or taken away. For example, a high entropy attribute will convey a high initial uncertainty about its actual value. If we then learn the value, we have acquired an amount of information equal to the initial uncertainty, i.e. the entropy we had originally about the value. Thus, information and entropy are two sides of the same coin. In this paper, we propose to apply entropy to a data set in a controlled way. This operation will take away data utility from the user but will in exchange generate protection. As such, data utility and protection also become two sides of the same coin, albeit in that case they are inversely related.

In information theory, a basic way to capture uncertainty is majorization [12]. Assume two vectors $x = (x_1, ..., x_N)^T$ and $y = (y_1, ..., y_N)^T$ that represent probability distributions, with the elements of each vector pre-ordered in decreasing order. The vector $x$ is said to majorize $y$, usually noted as $x \succ y$, if and only if the largest element of $x$ is greater than the largest element of $y$, the largest two elements of $x$ are greater than the largest two elements of $y$, and so on… [10]. Equivalently, that means that the probability distribution represented by $x$ is more narrowly peaked than $y$, in turn implying that $x$ conveys less uncertainty than $y$, thus that $x$ has less entropy than $y$.

In the privacy literature there is no such well-defined notion of information and no associated concepts such as majorization. What is meant as information for the meaningful exploitation of a data set lies in the eye of the user. For example, one user may be interested in the ability to perform some simple statistical requests such as cross-tabulations and thus will call information the analytical validity of such requests on anonymized data and their close proximity with the same requests performed on the original data set. Another user may be only interested in the ability to perform some econometric analyses, and thus again will qualify an anonymized data as informative given, for example, the validity of some OLS outputs made on it. Of course, and because the needs of users can be quasi-infinitely rich, one is left with a severe problem of diversity for evaluating the information content of an anonymized data set. In the paper, we reasonably assume that the original data set always provides the highest utility and analytical value to the user, and thus that an anonymized data set always entails a loss of utility.

**Acknowledgements.** Partial funding from the European Commission under project H2020-871042 "SoBigData++" is acknowledged. The second author is also partially funded by an ICREA Acadèmia Prize.